\documentclass[journal=nalefd,manuscript=article]{achemso}

\usepackage{chemformula} 
\usepackage[T1]{fontenc} 

\usepackage{ulem}
\usepackage{xcolor}
\usepackage{fancyhdr,graphicx}
\usepackage{bm}
\usepackage{color}
\usepackage{soul}
\usepackage{amsfonts}
\usepackage{times}
\usepackage{amsmath}
\usepackage{parskip} 
\usepackage{natbib}
\usepackage{chemformula} 
\usepackage[colorlinks=true, citecolor=red]{hyperref}
\setcitestyle{journalcolor= blue}
\usepackage[T1]{fontenc} 
\usepackage{wrapfig}
\usepackage{amsmath} 
\newcommand{\angstrom}{\textup{\AA}}
\usepackage{easyReview}

\usepackage{babel}

\title{Observation of time-reversal symmetric Hall effect in graphene-\ch{WSe2} heterostructures at room temperature}
\author{Priya Tiwari$^{1,2}$$^\dagger$, Divya Sahani$^1$$^\dagger$, Atasi Chakraborty$^{3,4}$, Kamal Das$^3$, Kenji Watanabe$^5$, Takashi Taniguchi$^6$, Amit Agarwal$^{3*}$, and Aveek Bid$^1$}
\email{amitag@iitk.ac.in, aveek@iisc.ac.in}
\affiliation{$^1$Department of Physics, Indian Institute of Science, Bangalore 560012, India \\
	$^2$ Braun Center for Submicron Research, Department of Condensed Matter Physics, Weizmann Institute of Science, Rehovot, Israel\\
	$^3$ Department of Physics, Indian Institute of Technology Kanpur, Kanpur-208016, India\\
	$^4$ Institute of Physics,  Johannes Gutenberg Universit\"{a}t, Staudinger Weg 7, 55128 Mainz, Germany \\
	$^5$ Research Center for Functional Materials, National Institute for Materials Science, 1-1 Namiki, Tsukuba 305-0044, Japan \\
	$^6$ International Center for Materials Nanoarchitectonics, National Institute for Materials Science, 1-1 Namiki, Tsukuba 305-0044, Japan \\
	$^\dagger$ These authors contributed equally.}
\begin{document}
	\begin{abstract}
		In this letter, we provide experimental evidence of the time-reversal symmetric Hall effect in a mesoscopic system, namely high-mobility graphene/\ch{WSe2} heterostructures. This linear, dissipative Hall effect, whose sign depends on the sign of the charge carriers, persists up to room temperature. The magnitude and the sign of the Hall signal can be tuned using an external perpendicular electric field. Our joint experimental and theoretical study establishes that the strain induced by lattice mismatch, or angle inhomogeneity, produces anisotropic bands in graphene while simultaneously breaking the inversion symmetry. The band anisotropy and reduced spatial symmetry lead to the appearance of a  time-reversal symmetric Hall effect. Our study establishes graphene-transition metal dichalcogenide-based heterostructures as an excellent platform for studying the effects of broken symmetry on the physical  properties of band-engineered two-dimensional systems.
	\end{abstract}

	\section{Introduction}

	Topological and band geometric effects in two-dimensional systems have attracted significant attention due to their fascinating physics and potential applications in spintronics and novel electronic devices~\cite{xiao_RMP2010_berry, Ahn2021, Gao2021, Bhalla_2022,spinreview}. Graphene-based heterostructures offer one such exciting platform for studying band geometric effects. The coupling to the charge, spin, and valley degrees of freedom in graphene lead to, among other things, a multitude of unconventional Hall effects; some prominent examples include the spin Hall~\cite{sinova2015spin, hirsch1999spin, bernevig2006quantumPRL, Tiwari2022}, the valley Hall~\cite{Xiao2012, cresti2016charge,mak2014valley, lee2016electrical,liu2019quantum,qiao2010quantum}, the nonlinear anomalous Hall~\cite{Shimazaki2015, Sui2015, Wellbank2016,sodemann_PRL2015_quantum,du_NC2019_dis,Sinha2021,Chakraborty_2022} and the layer contrasted Hall effect~\cite{Gao2021,zhai_arxiv2022_layer}. All these phases require a finite Berry curvature and  opposite anomalous velocity in the two valleys of graphene to exist~\cite{Shimazaki2015, Sui2015, Wellbank2016}. They fall under the category of dissipationless anti-symmetric Hall effect, in which the current always flows perpendicular to the electric field, and the polarity of voltage changes if one interchanges the source and drain contacts. What is under-appreciated is that a time-reversal symmetric system with anisotropic band velocity can also host non-trivial and symmetric Hall effects irrespective of the presence or absence of a finite Berry curvature. The focus of this Letter is the observation of a  finite `symmetric Hall response'~\cite{ho2021hall, nandy2017chiral, kumar2018planar}. The symmetric Hall effect can appear without time-reversal symmetry breaking and is unrelated to the so-called  quantum geometry of the wave function.

	In this letter, we report the observation of a large time-reversal symmetric, linear Hall response $\mathrm{R_{xy}(B=0)}$ originating from asymmetric momentum space band dispersion of the heterostructure. We find that the $\mathrm{R_{xy}(B=0)}$ flips sign as the Fermi level moves from the valence to the conduction band (mimicking the finite $\mathrm{B}$-field classical Hall signal in graphene) and is observed up to room temperature. We explain our observations using a realistic two-band model for an anisotropic band dispersion with an orthogonal anisotropy orientation in the two bands. We show that such band anisotropy can appear in graphene-\ch{WSe2} heterostructures through realistic model calculations.

	Single-layer \ch{WSe2} used as a substrate influences the graphene bands in two significant ways. The first of these is well studied:  Graphene on \ch{WSe2} possesses spin-split bands owing to the Ising-like SOC, which gives rise to the spin Hall effect~\cite{avsar2014spin, ghiasi2019charge,doi:10.1021/acsnano.0c07524}. The second effect, equally vital for our purposes but ill-explored to date, is the appearance of a substantial lateral strain in the graphene layer. We show that the effect of this proximity-induced SOC, lattice mismatch, and small angle rotation can induce gap and asymmetric dispersion in graphene, leading to the appearance of the symmetric Hall signal near the Dirac point. We verify that the symmetric, charge-carrier sign-dependent Hall response is absent in graphene devices without   the \ch{WSe2} layer. Note that previous studies on the SLG-\ch{WSe2} heterostructure (or graphene on transition metal dichalcogenides in general) focused primarily on the spin aspects of the  transport~\cite{avsar2014spin,ghiasi2019charge,herling2020gate,dastgeer2022gate, PhysRevB.106.165420} where a non-local signal is measured as a signature of the spin Hall effect and weak (anti-) localization measurements were used to quantify the spin-orbit coupling strength~\cite{Wang2015,wang2016origin,tobias2017magneto, Wakamura2019spin,fulop2021boost, PhysRevLett.126.096801,doi:10.1021/acsnano.0c07524, tiwari2021electric}. Interestingly, these studies did not probe the finite Hall effect without a magnetic field. This makes our observation of the time-reversal symmetric non-trivial Hall effect in this system unique.

	\section{Results}

	\subsection{Device characteristics}

\begin{figure*}[t!]
		\begin{center}
			\includegraphics[width=\columnwidth]{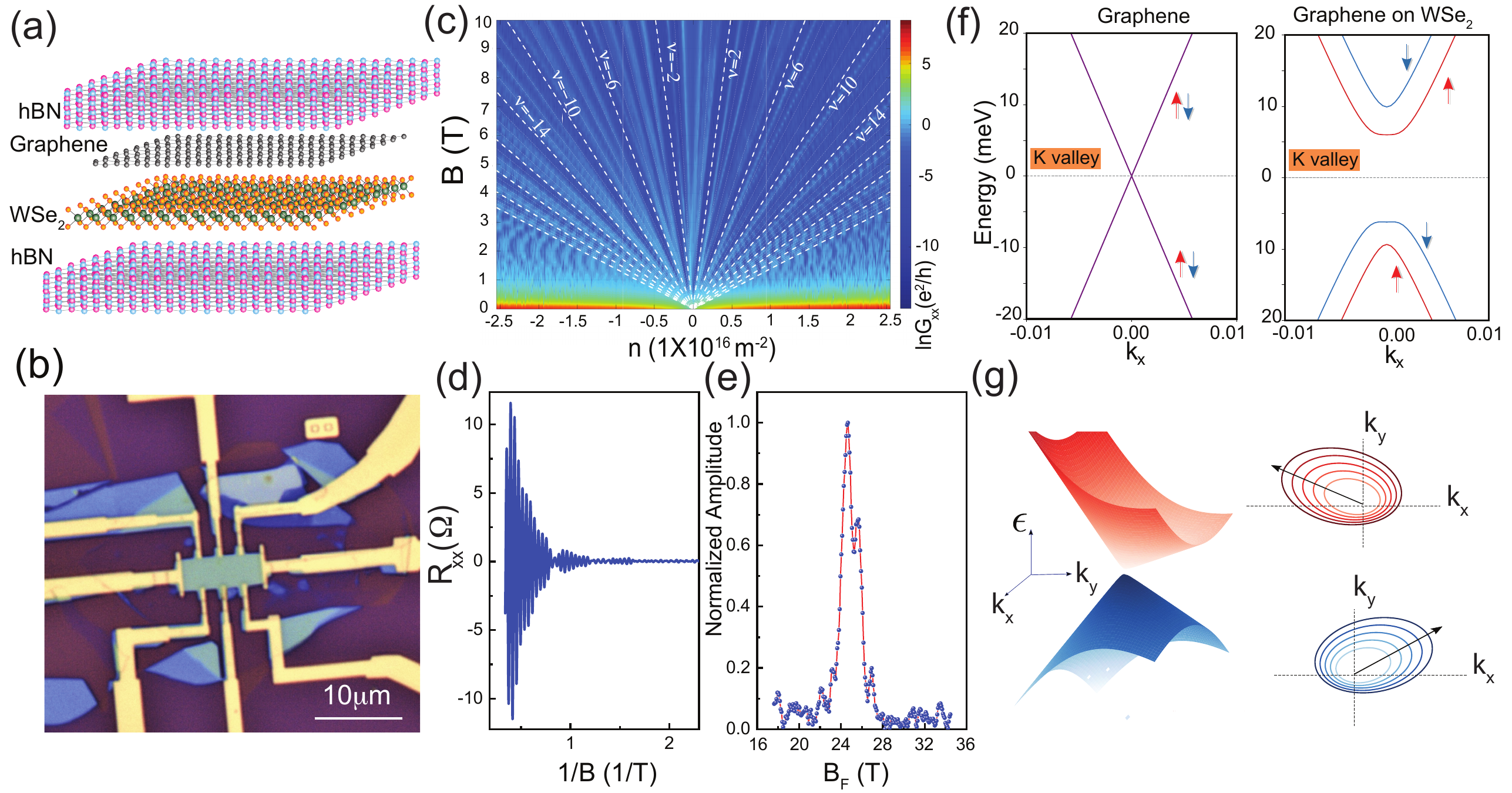}
			\small{\caption{\textbf{Device characteristics and band dispersion:} (a) Schematic of the graphene/\ch{WSe2} layers encapsulated in hBN illustrating the sequence of crystal stacking. (b) Optical image of the device. (c) Map of the longitudinal conductance ($\mathrm{G_{xx}(B)}$)  with varying carrier density $n$ and perpendicular magnetic field $\mathrm{B}$ at $\mathrm{T}\sim20$~mK. The thicker dashed lines correspond to the signature plateaus of single-layer graphene. Thinner lines mark the broken-symmetry phases indicating complete lifting of the spin and valley degeneracies at low-$\mathrm{B}$. (d) SdH oscillations versus $\mathrm{1/B}$ at $V_{\rm bg}=-40$~V. (e) Fourier spectrum of the SdH oscillations: two peaks are distinctly visible, establishing the presence of two Fermi surfaces.  (f) Schematic of the band dispersion of the $K$ valley of monolayer graphene (left panel) and graphene on \ch{WSe2} heterostructure (right panel). The \ch{WSe2} layer essentially lifts the spin degeneracy of the low-lying energy bands and opens up a gap at the Fermi energy. (g) Schematic of the band structure of strained graphene on \ch{WSe2} with the orientation of band anisotropy being orthogonal in the conduction and valence bands. This is highlighted by the arrows in the contour plots. }
				\label{fig:figure1}}
		\end{center}
	\end{figure*}

	Heterostructures of single-layer graphene (SLG) and single-layer \ch{WSe2}, encapsulated by crystalline hexagonal boron nitrate (hBN), were fabricated using a dry transfer technique~\cite{pizzocchero2016hot,wang2013one}. One-dimensional electrical contacts were formed by electron beam lithography, followed by etching (using a mixture of \ch{CHF3} and \ch{O2}) and deposition of 5~nm/60~nm Cr/Au contacts. A separate deposition was done for the top-gate electrode. See the Supporting Information (SI) for details. A schematic of the device structure is shown in Fig.~\ref{fig:figure1}(a), and an optical image of the device is shown in Fig.~\ref{fig:figure1}(b). The dual-gated architecture of the devices allows independent control of the charge-carrier density $n$ and the vertical displacement field $\mathrm{D}/\epsilon_{0}$; $n\mathrm{=(C_{tg}V_{tg} + C_{bg}V_{bg})/e} -n_{0}$ and $\mathrm{D=(C_{tg}V_{tg} - C_{bg}V_{bg})/2-D_{0}}$. Here $\mathrm{C_{bg}}$ ($\mathrm{C_{tg}}$) is the capacitance per unit area of the back-gate (top-gate), $\mathrm{V_{bg}}$ ($\mathrm{V_{tg}}$) is the back-gate (top-gate) bias. $n_0$ and $\mathrm{D_0}$ are the residual charge carrier density and residual vertical displacement field induced by impurities in the device channel.

	Electrical transport measurements were performed at 10~nA source-drain current using low-frequency lock-in detection techniques. All data were obtained at 20~mK unless specified otherwise. The measurements were performed on multiple devices; the results were similar. In the main manuscript, we present the data from a single device, SW1, unless specified. The data from three other devices (SW2, SG3, SG4) are presented  in Fig.~S5, S6, and S7 of the SI.

	A map of the measured longitudinal conductance $\mathrm{G_{xx}}$ as a function of charge carrier density $n$ and perpendicular magnetic field $\mathrm{B}$ is shown in Fig.~\ref{fig:figure1}(c). The appearance of broken symmetry quantum Hall states at low $\mathrm{B}$-fields implies a complete lifting of the spin and valley degeneracies in SLG bands. The splitting of the spin-degenerate bands in SLG (shown schematically in Fig.~\ref{fig:figure1}(f)) is also evident from the beating pattern seen in the Shubnikov de Haas oscillations [Fig.~\ref{fig:figure1}(d)], and the double periodicity in the corresponding Fourier spectrum [Fig.~\ref{fig:figure1}(e)]. Fig.~\ref{fig:figure1}(g) is a schematic representation of the band structure of strained graphene on \ch{WSe2} with the orientation of the band anisotropy in the valence and conduction band being orthogonal (see Sec. I of SI for more details). The lifting of spin-and valley degeneracies in the band dispersion (along with the high field-effect mobility $\mathrm{\mu} \sim 140,000~ \mathrm{cm^2 V^{-1} s^{-1}}$ of the device)  shows that the graphene and \ch{WSe2} interface is atomically clean with significant interfacial coupling and minimal random potential fluctuations.

	\subsection{Room temperature symmetric Hall effect at $\mathbf{\mathrm{B=0}}$~T}

\begin{figure*}[t!]
		\begin{center}
			\includegraphics[width=\columnwidth]{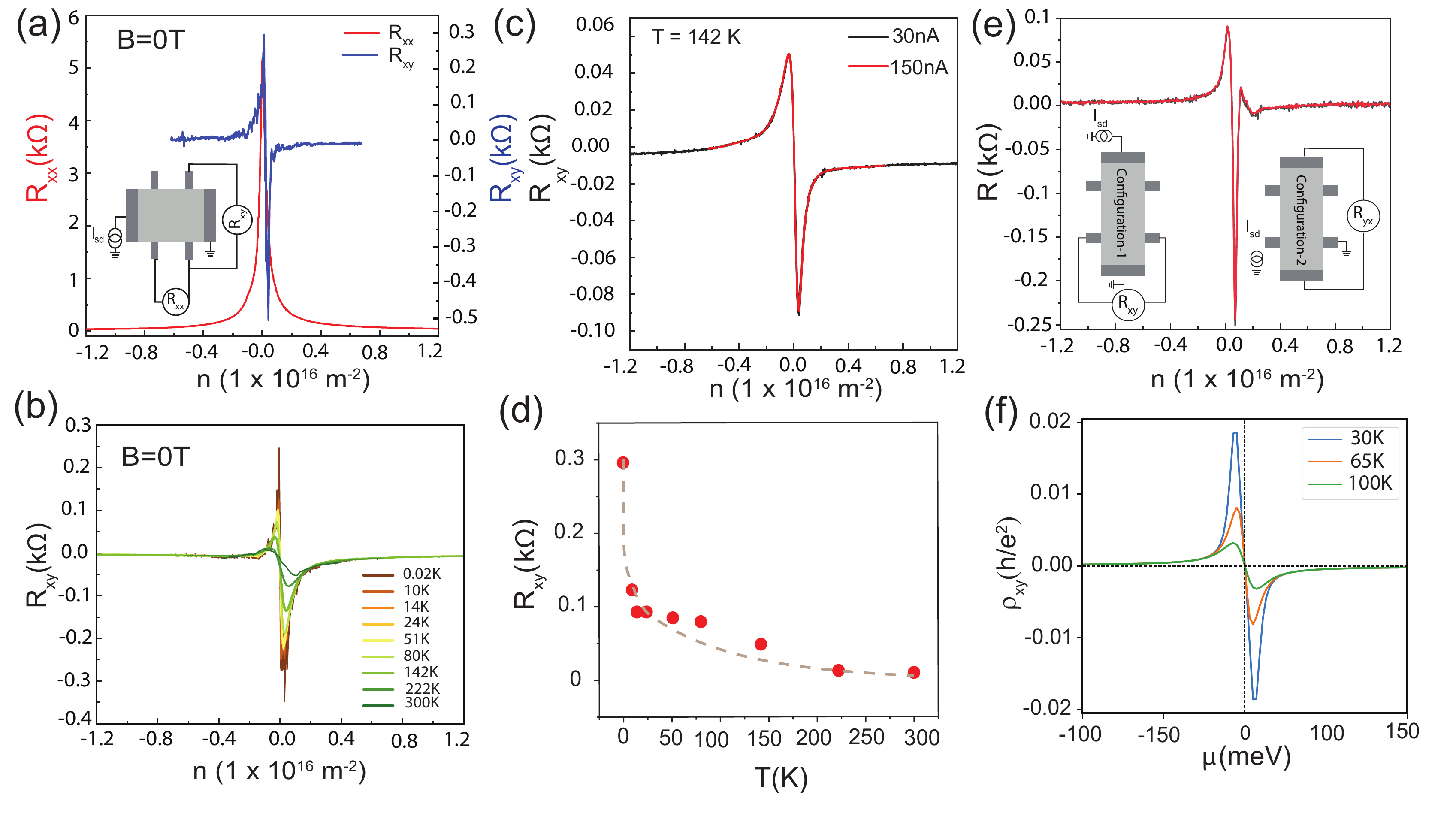}
			\small{\caption{\textbf{Sign reversing Symmetric Hall effect} (a) Plots of the zero magnetic-field longitudinal resistance $\mathrm{R_{xx}(B=0)}$ (left-axis, red line) and zero magnetic-field transverse resistance $\mathrm{R_{xy}(B=0)}$ (right-axis, blue line) versus $n$; the data were measured at $T=20$~mK.   (b) $\mathrm{R_{xy}(B=0)}$ response as a function of $n$ at few representative values of temperature; the AHE persists up to 300~K. (c) Plot of $\mathrm{R_{xy}(B=0)}$ as a function of $n$ for two different values of electrical current; the data were taken at $T=142$~K. (d) Plot of the peak value of $\mathrm{R_{xy}(B=0)}$ versus $\mathrm{T}$. The dotted line is a guide to the eye.
					(e) Plot of transverse resistance versus number density in two different configurations for
					device SW2. Configuration 1 measures $\mathrm{R_{xy}(B=0)}$ and configuration 2 measures $\mathrm{R_{yx}(B=0)}$. One can see that $\mathrm{R_{xy}(B=0)} = \mathrm{R_{yx}(B=0)}$; the measured Hall effect is symmetric on interchanging the voltage and current leads. (f)  Theoretically calculated symmetric Hall resistivity ($\rho_{xy}$) for three different temperatures.
				} 	\label{fig:figure2}}
		\end{center}
	\end{figure*}

	In Fig.~\ref{fig:figure2}(a), we present the data for the longitudinal resistance, $\mathrm{R_{xx}}$ (left-axis, red line), and transverse resistance, $\mathrm{R_{xy}}$ (right-axis, blue line)  measured at $\mathrm{B=0}$~T. We observe a finite Hall signal in a narrow range of charge carrier densities $\Delta n = \pm 10^{15}$~$\mathrm{m^{-2}}$ centered about the charge neutrality point.  $\mathrm{R_{xy}(B=0)}$ measured for other contact configurations give quantitatively similar results (see Fig.~S5 in SI). Strikingly, the $\mathrm{R_{xy}(B=0)}$ features a change in the sign about the charge neutrality point -- it is positive for $n < 0$ (hole-band) and negative for $n>0$ (electron-band). The current independence of $\mathrm{R_{xy}(B=0)}$ establishes it to be a linear Hall effect [see Fig.~\ref{fig:figure2}(c)].  The finite  Hall response survives at least to room temperature with diminished amplitude as shown in Figs.~\ref{fig:figure2}(b) and (d). Importantly, the transverse signal is symmetric upon exchange of the voltage and current leads; $\mathrm{R_{xy}(B=0)}=\mathrm{R_{yx}(B=0)}$ [see Fig.~\ref{fig:figure2}(e)]. This observation in hBN/graphene/\ch{WSe2}/hBN heterostructures of a room temperature, $\mathrm{B=0}$ symmetric Hall effect that changes sign across the charge neutrality point is unique, and it is the central result of this letter.

	A finite zero-magnetic-field Hall response is often seen in graphene devices near the charge neutrality point (see Fig.~S7 in SI). This can originate from trivial effects that mix the longitudinal and transverse signals (e.g., misaligned probes or misalignment of the measurement axis with the crystal axis); thus, the sign of this signal is independent of the  charge carrier type. We emphasize that our observation is distinct from this trivial case in one very important aspect: In graphene/\ch{WSe2} heterostructures, the hall response flips sign as the carrier type changes from electrons to holes (see SI for a comparison of  data in graphene and graphene/\ch{WSe2} devices).  The data for $\mathrm{R_{xy}(B=0)}$ were also reproduced in cryostats without a superconducting magnet, ruling out the remnant field of a magnet as the origin of the observed Hall response.

	Having ruled out experimental artifacts as the origin of our unusual observation, we develop below a detailed model that explains our essential observations. Asymmetric Hall response can arise from the Drude mechanism in materials with an anisotropic band dispersion. The symmetric Hall conductivity is specified by as\cite{Ashcroft76,ho2021hall} $\sigma_{xy}=e^2 \tau \int_{\bm k} v_x v_y (-f')$. Here, $v_i = \partial_{k_i} \epsilon $ is the band velocity, $f'$ is the energy derivative of the Fermi distribution function, and $\tau$ is the scattering time. However, such a symmetric Hall response usually does not change its sign when going from the valence to the conduction band, contrary to our experimental observations (see Fig. S6 in the SI).
	We argue below that such sign reversal can arise in particle-hole symmetry broken systems with different anisotropy orientations in the electronic dispersion of the conduction and valence band.

	To gain a complete understanding of the unusual Hall effect observed in SLG/\ch{WSe2}, we start by noting that the band structure of this system is anisotropic.
	This is schematically highlighted in Fig.~\ref{fig:figure1}(g). The anisotropy axis in the conduction and the valence bands is close to orthogonal (see Fig.~S1 of the SI for realistic calculations for graphene-WSe$_2$ heterostructure). This band structure is reminiscent of that seen in periodically strained bilayer graphene that hosts a time-reversal symmetric linear pseudo-planar Hall effect~\cite{ho2021hall}. Such an anisotropic band dispersion preserves time-reversal symmetry but breaks particle-hole symmetry (see SI for details). Let us consider a symmetric system in which we add a small anisotropy. The anisotropy  modifies the band structure to induce a change in the band velocity,  $v_{x/y} \to v_{x/y}^{\rm symm} + \delta v_{x/y}$. 	In a simplistic scenario where $\delta v_{x/y}$ are independent of the Bloch momentum, we have, $\sigma_{xy} \propto \delta v_x \delta v_y |\mu|$. One can show that if the tilt direction of the conduction and the valance band are orthogonal to each other, as it is for SLG/\ch{WSe2} [see Fig.~\ref{fig:figure1}(e)], the product $\delta v_{x} \delta v_{y}$ and consequently the symmetric Hall response reverses its sign as we go from conduction to valence band.
	This orthogonal orientation of the band anisotropy in the conduction and the valence band is the origin of the sign reversing symmetric Hall effect. See SI for the generic model of sign reversing symmetric Hall effect and detailed calculations for the graphene-WSe$_2$ heterostructure.  Figure~2(f) presents the symmetric Hall conductivity for an effective two-band model, which qualitatively captures the sign-reversing symmetric Hall effect (also see Fig.~S2 of the SI).

	We do not find evidence of sign reversing symmetric Hall effect in hBN/graphene/hBN devices without the intervening \ch{WSe2} layer (SI). Thus, the orthogonal asymmetry in the conduction and valence band induced by \ch{WSe2}-graphene combination is essential for this effect.

	The band anisotropy in SLG on \ch{WSe2} can be traced back to lattice mismatch-induced strain. The lattice constant of graphene is $\sim2.46$~$\angstrom$ while that of \ch{WSe2} is $\sim3.27$~$\angstrom$. This large lattice mismatch generates a significant strain across the graphene flake as the heterostructure relaxes to the stable ground state. From Raman spectroscopy, we estimate the magnitude of the strain on the SLG layer in our hBN/SLG/\ch{WSe2}/hBN heterostructure to be $\approx 0.15\%-0.20\%$ (see SI ). We find that the strain magnitude decreases with temperature (see SI for details). Our observed symmetric Hall effect survives the combined action of i) decreasing strain and band anisotropy with increasing temperature and ii) the thermal broadening of the Fermi function to be finite at room temperature.

	\subsection{Symmetric Hall response with vertical displacement and magnetic field}

\begin{figure}[t!]
		\includegraphics[width=\columnwidth]{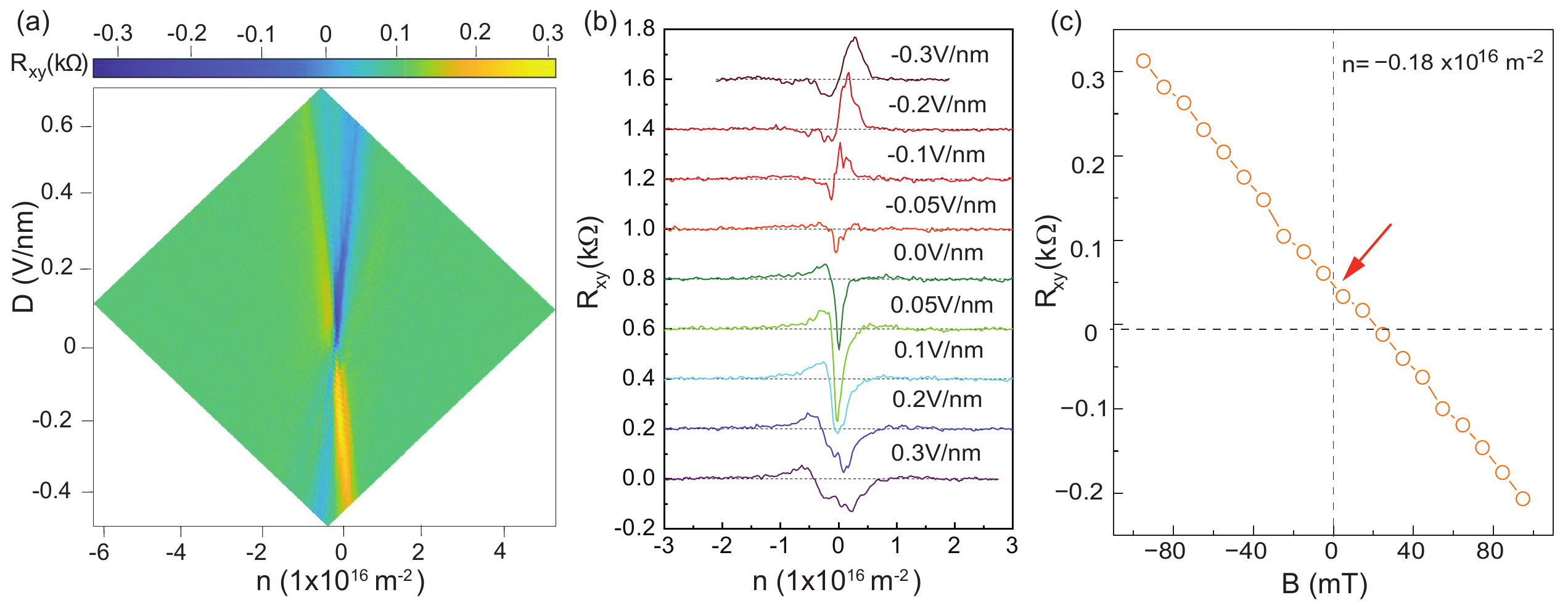}
		\small{\caption{\textbf{Dependence of the transverse resistance $\mathrm{R_{xy}}$ on $\mathrm{D}$ and $\mathrm{B}$.} (a) A 2-dimensional contour map of $\mathrm{R_{xy}(B=0)}$ plotted in the $n-\mathrm{D}$ plane.  (b) Plots of $\mathrm{R_{xy}(B=0)}$  versus  $n$ for different values of $\mathrm{D}$. The data have been vertically shifted by 200~$\Omega$ for clarity. The dashed horizontal line for each plot marks the zero of $\mathrm{R_{xy}(B=0)}$. (c) A representative plot of $\mathrm{R_{xy}}$ versus $\mathrm{B}$ measured at $n=-0.18\times 10^{16}$~$\mathrm{m^{-2}}$, an arrow marks the value of the $B=0$ Hall resistance.   }
			\label{fig:figure4}}
	\end{figure}

	Having demonstrated the symmetric Hall effect, we now focus on the dependence of the symmetric Hall response on a perpendicular displacement field $\mathrm{D}$  (Fig.~\ref{fig:figure4}). It is illuminating to map the transverse zero-$\mathrm{B}$-field conductivity $\mathrm{R_{xy}(B=0)}$ data in the  $n - \mathrm{D}$ plane (Fig.~\ref{fig:figure4}(a)). The plot shows $\mathrm{R_{xy}(B=0)}$ to be finite only at the band edges, consistent with the idea that the longitudinal conductivity is large inside the band and vanishingly small in the vicinity of the band edges. 	This can be seen clearly in the line plots of  $\mathrm{R_{xy}(B=0)}$ for different values of $\mathrm{D}$ shown in Fig.~\ref{fig:figure4}(b). Note that the plots are vertically offset by 200~$\Omega$ for clarity. The measured $\mathrm{R_{xy}(B=0)}$ has an intriguing $\mathrm{D}$ dependence; it changes its sign as the direction of $\mathrm{D}$ flips [Figs.~\ref{fig:figure4}(a-b)].

	Measurements in a finite magnetic field $\mathrm{B}$ applied perpendicular to the device interface (see SI) reveal the interplay between the classical Hall effect and the $\mathrm{B=0}$ symmetric Hall effect. The data smoothly crosses over from the symmetric Hall phase at $\mathrm{B=0}$ to the conventional Hall phase at finite $\mathrm{B}$-field with an anti-crossing feature. This feature resembles the planar Hall effect in corrugated bilayer graphene~\cite{ho2021hall}. A non-zero intercept of the plot of $\mathrm{R_{xy}}$ versus $\mathrm{B}$ [shown for a fixed $n$ in Fig.~\ref{fig:figure4}(c)] on the $\mathrm{B}$-axis captures the symmetric Hall effect. We note that $\mathrm{R_{xy}}$ is non-hsyteretic in the presence of a small non-quantizing magnetic field (see Fig. S6 (d) of SI), ruling out emergent ferromagnetism in the system.

\begin{figure*}[t!]
		\begin{center}
			\includegraphics[width=\columnwidth]{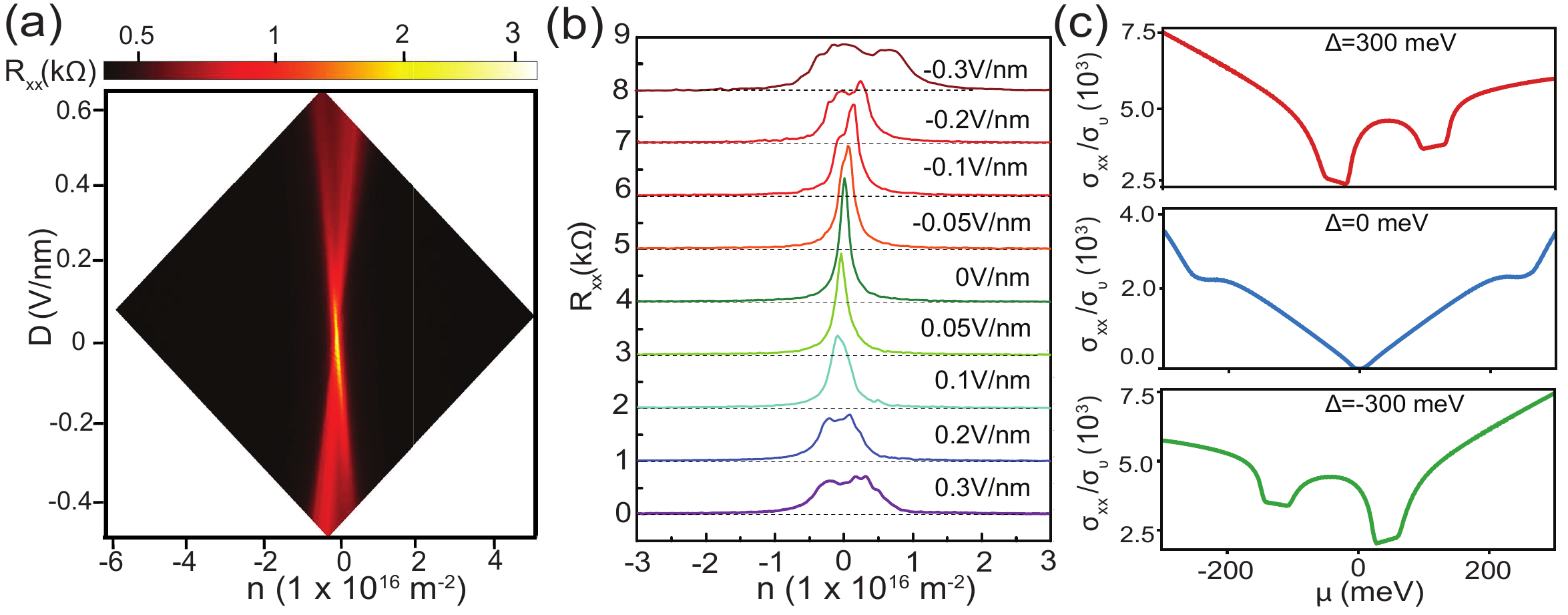}
			\small{\caption{\textbf{Dependence of $\mathrm{R_{xx}(B=0)}$ on $\mathrm{D}$}.  (a) A 2-dimensional contour map of $\mathrm{R_{xx}(B=0)}$ plotted in the $n-\mathrm{D}$ plane. (b) Plots of $\mathrm{R_{xx}(B=0)}$  versus  $n$ for different values of $\mathrm{D}$. The data have been vertically shifted by 1~k$\Omega$ for clarity. The dashed horizontal line for each plot is the zero of the y-axis. (c) Variation of the calculated Drude conductivity $\sigma_{xx}$ with energy ($\mu$) for three different values of the interlayer potential induced by the applied electric field, $\Delta=300$~meV (red line), 0 meV (blue line) and $-300$ meV (green line), respectively. The values of $\sigma_{xx}$  have been scaled by $\sigma_v$ where $\sigma_v={e^2 \tau}/{4\pi^2\hbar^2}$.
				}
				\label{fig:figure5}}
		\end{center}
	\end{figure*}

	In Fig.~\ref{fig:figure5}(a), we present a plot of $\mathrm{R_{xx}}$ in the $n-\mathrm{D}$ plane measured at $\mathrm{B=0}$. We observe that with increasing $\mathrm{D}$, the resistance peak at the charge neutrality point splits into two maxima. This feature can be better appreciated from Fig.~\ref{fig:figure5}(b), where we show individual plots of $\mathrm{R_{xx}(B=0)}$ versus $n$ at several representative values of $\mathrm{D}$. At higher values of $|\mathrm{D}|$, we find two distinct peaks in $\mathrm{R_{xx}}$ separated by a shallow valley. Such a displacement field-dependent dispersion of the bands near the Dirac point is not captured by the existing models for graphene/\ch{WSe2} heterostructures~\cite{Wang2015,gmitra2016trivial,offidani2017optimal,cummings2017giant,garcia2018spin,li2019twist,zubair2020influence,kumar2021zero}. To remedy this, we construct a new model Hamiltonian for the graphene/\ch{WSe2} system, retaining both the \ch{WSe2} and the graphene Hamiltonian blocks, which allows us to include the impact of a vertical displacement field systematically (see SI for details).  Fig.~\ref{fig:figure5}(c) is a plot of the theoretically calculated $\mathrm{\sigma_{xx}}$ as a function of the chemical potential -- the panels show the splitting of the conductivity minima into two asymmetric conductivity minima  at finite $\mathrm{D}$. Our model thus reproduces the prominent features of $\mathrm{\sigma_{xx}}$ both at zero displacement field~\cite{gmitra2016trivial,cummings2017giant} and at a finite $\mathrm{D}$, along with the observed symmetric Hall effect.

	\section{Discussion}

	To summarize, we report the first observation of room temperature time-reversal symmetric Hall effect in high-mobility heterostructures of graphene/\ch{WSe2}. Primarily known for their promising spintronics aspects, the charge Hall response of such a heterostructure was expected to be relatively mundane. Contrary to this, we measure a time-reversal symmetric Hall effect at zero magnetic field. More interesting, the observed symmetric Hall effect changes sign across the charge neutrality point and persists till room temperature. We show that the symmetric Hall effect arises from the anisotropy of the electronic band structure. The sign reversal of the observed Hall effect arises from the different orientations of the anisotropy direction in the conduction and the valence band. 	The symmetric Hall response features a unique perpendicular electric field tunability. Our work establishes graphene-\ch{WSe2} heterostructure as an excellent platform for further exploration of the interplay of charge, spin, and valley responses in band-engineered two-dimensional systems.

	\section{AUTHOR INFORMATION}

	\textbf{Author Contributions}

	A.B., P.T., and D.S. conceptualized the study, performed the measurements, and analyzed the data.   A.A., A.C., and K.D. performed the theoretical analysis. K.W. and T.T. grew the hBN single crystals. All the authors contributed to preparing the manuscript.

	\textbf{Notes}

	The authors declare no competing financial interest.

	\begin{acknowledgement}
		The authors acknowledge stimulating discussions with Sumanta Tewari and Jay Deep Sau.	A.B. acknowledges funding from the Department of Science \& Technology FIST program, DST fellowship (No. DST/SJF/PSA01/2016-17), and the U.S. Army DEVCOM Indo-Pacific (Project number: FA5209   22P0166). A.C. acknowledges the Indian Institute of Technology, Kanpur, and the Science and Engineering Research Board (SERB) National Postdoctoral Fellowship (PDF/ 2021/ 000346), India, for partial financial support. A.C. acknowledges Alexander Von Humboldt Post-doctoral fellowship, Germany, for funding. A.A. acknowledges the Science and Engineering Research Board for Project No. MTR/2019/001520, and the Department of Science and Technology for Project No. DST/NM/TUE/QM-6/2019(G)-IIT Kanpur of the Government of India for funding. K.W. and T.T. acknowledge support from JSPS KAKENHI (Grant Numbers 19H05790, 20H00354, and 21H05233).
	\end{acknowledgement}

	\begin{suppinfo}

		Supporting information contains detailed discussions of (a) Model Hamiltonian of Graphene/\ch{WSe2} heterostructure, (b) Generic low energy model for sign reversing symmetric Hall effect, (c) Drude longitudinal conductivity, (d) Device fabrication, (e) Raman shift and strain (e) Additional Data for symmetric Hall effect on device SW1. (f) Additional Data for symmetric Hall effect on device SW2 (g) Transverse resistance Data on hBN/graphene/hBN device at B=0.

	\end{suppinfo}
	\clearpage


	\clearpage

	\clearpage

\section*{Supplementary Information}
	\section{I. Model Hamiltonian of Graphene \ch{WSe2} heterostructure}
\noindent In this section, we construct the low energy model Hamiltonian of monolayer graphene on a \ch{WSe2} layer. Going beyond the effective graphene model as reported in recent literature~\cite{PhysRevB.92.155403,gmitra2016trivial,cummings2017giant}, we explicitly solve for the composite low energy Hamiltonian for the graphene-\ch{WSe2} heterostructure to capture the effect of perpendicular electric field correctly. We solve the following low-energy Hamiltonian
\begin{equation}
	H_{tot}
	=
	\begin{pmatrix}
		H_{k}^{g} & H_{t}\\
		H_{t}^\dagger & H_{tot}^{ws}
	\end{pmatrix}
	+ H_\perp
\end{equation}
Here, $H_k^{g}$ and $H_{tot}^{ws}$ are the onsite Hamiltonian for graphene and the \ch{WSe2} respectively. The interaction between graphene and \ch{WSe2} layer has been included through spin and valley conserved off-diagonal hopping ($H_t$). The effect of the perpendicular electric field is captured through the diagonal matrix $H_\perp$.


\begin{figure*}[t]
	\begin{center}
		\includegraphics[width=0.85\columnwidth]{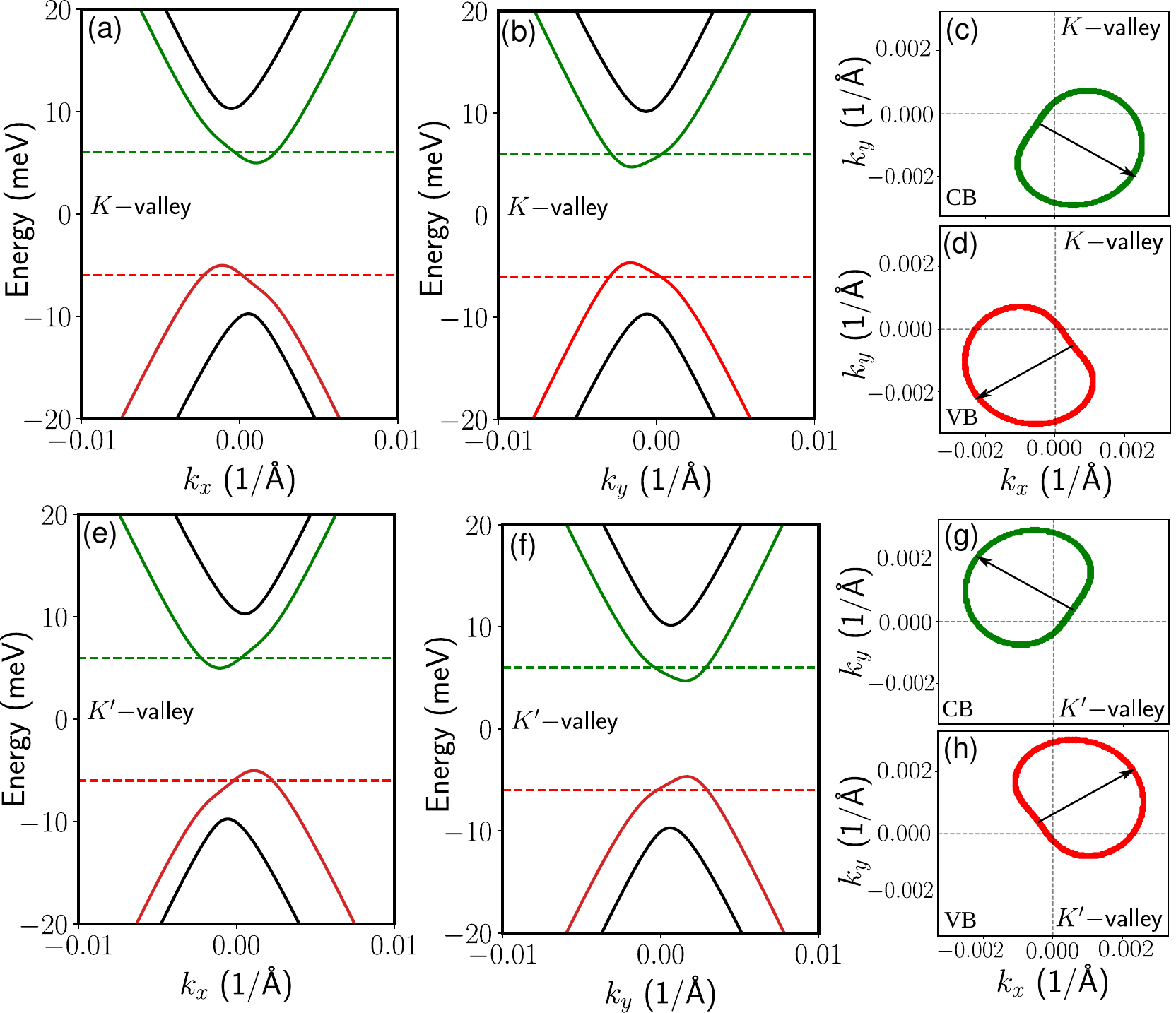}
		\caption{\textbf{Anisotropy of graphene-\ch{WSe2} band structure in the conduction and valence band in both  valleys.} (a-b) and (e-f) show the energy dispersion of graphene-\ch{WSe2} composite Hamiltonian, in presence of an off-diagonal hopping, along $k_x~(k_y)$ direction for $K-$valley and $K^\prime$-valley respectively. The valley-specific constant energy contours of the valence band (VB) and conduction band (CB) in the vicinity of the band edges ($E = \pm 6$ meV)  are shown in (c-d) and (g-h). The opposite orientation of the anisotropy axis in the two valleys ensures the presence of time reversal symmetry. This was also explicitly checked via the relation $E_k^{K}= E_{-k}^{K^\prime}$.
			The anisotropic band structure plays a crucial role in generating a symmetric Hall effect. The different (close to orthogonal) orientation of the anisotropy axis in the conduction and valance bands ensures that the symmetric Hall response changes sign as one goes from electron-doped to hole-doped region. \label{Fig:contour}}
	\end{center}
\end{figure*}
We consider the monolayer of \ch{WSe2} in the $x$-$y$ plane in the presence of intrinsic spin-orbit coupling (SOC) ($H_{sym}^{ws}$), spin Zeeman field ($\Delta_0^{ws}$). In addition, finite Rashba SOC term  ($H_R^{ws}$) is also considered within the \ch{WSe2} sector~\cite{Sattari2022E,Novakov2021I}. Including all these effects, the two-dimensional extended Dirac Hamiltonian ($H_{tot}^{ws}$) of \ch{WSe2} monolayer can be written as
\begin{figure*}[t]
	\begin{center}
		\includegraphics[width=0.4\columnwidth]{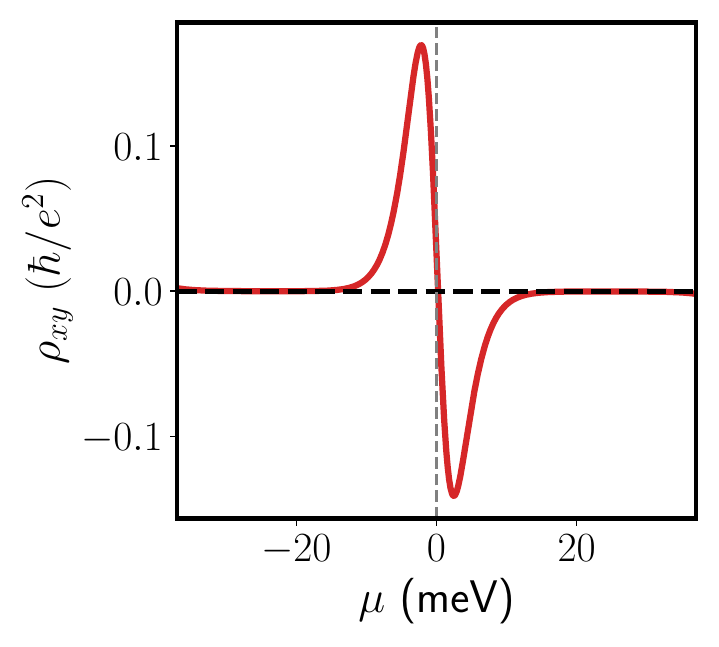}
		\caption{Drude Hall resistivity at $T=30$~K calculated from the graphene-\ch{WSe2} composite model in absence of perpendicular electric field. The Hall resistivity changes its sign from the valence ($\mu<0$) to the conduction ($\mu>0$) sector. The black dashed line is the calculated anomalous Hall resistivity of the system. The vanishing AHE indicates that the system preserves time-reversal symmetry.  \label{Fig:sigma}}
	\end{center}
\end{figure*}

\begin{equation}
	H_{tot}^{ws}=H_k^{ws}+H_{sym}^{ws}+H_R^{ws}.
\end{equation}

\begin{figure*}[t!]
	\includegraphics[width=\columnwidth]{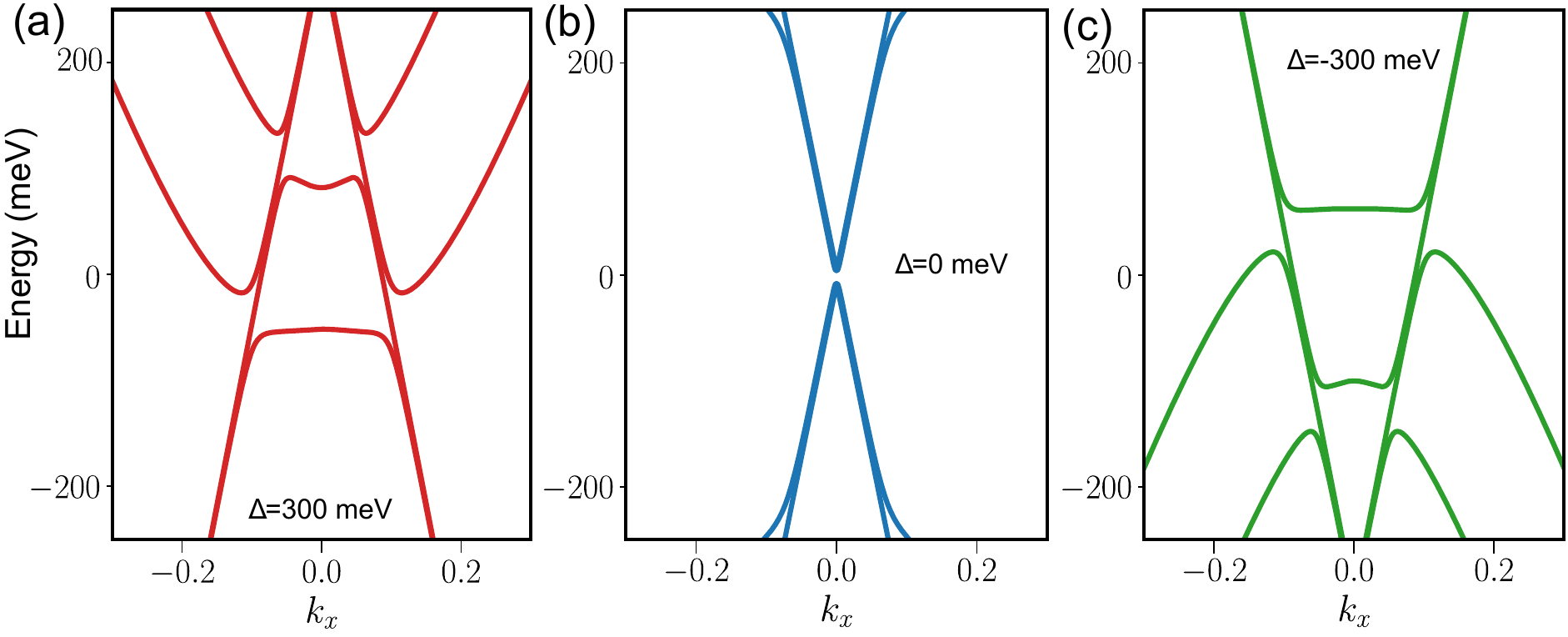}
	\small{\caption{Impact of the electric field on the band structure of graphene/\ch{WSe2} heterostructure. (a), (b) and (c) show the band dispersion in the presence of electric field values $\Delta=$ 300~meV, 0 meV, and -300~meV, respectively. The external electric field changes the low energy band dispersion of the composite graphene-\ch{WSe2} heterostructure,  inducing a  metal-insulator transition.}
		\label{fig:s1}}
\end{figure*}
The explicit forms of each term are as follows,
\begin{align}
	H_k^{ws}&= v_F^{ws}[\xi \sigma_xk_x+\sigma_yk_y] +\Delta_0^{ws}\sigma_z~, \nonumber \\
	H_{sym}&=\frac{1}{2}[\lambda_c(\sigma_z+\sigma_0)+\lambda_v(\sigma_z-\sigma_0)]~, \nonumber \\
	H_R^{ws}&= \lambda_R[\xi \sigma_x S_y-\sigma_y S_x]~,\label{Ham_term}
\end{align}
where $\xi=\pm 1$ for $K$ and $K'$ valley respectively.  As in the monolayer \ch{WSe2}, two degenerate but inequivalent valleys ($K$ and $K'$) are separated by a large momentum; we can split the total Hamiltonian into two valley-specific parts. Here, we have considered $v_F^{ws}\equiv$1.83 eV.\AA~as the Fermi velocity of \ch{WSe2}. $\Delta_0$ represents the mass term that breaks the inversion symmetry. Here, $\lambda_c$ and $\lambda_v$ correspond to the SOC strengths of conduction and valence bands. In general, the valence band  ($\lambda_v\sim112.5$ meV) of \ch{WSe2} possesses larger SOC strength than the conduction band ($\lambda_c\sim7.5$meV), promoting relatively larger splitting in the valence band~\cite{Tahir2016M, Tahir2018E}. For simplicity of the calculation, we choose the SOC strengths of both the conduction and valence bands to be equal, $\lambda_c=\lambda_v=$7.5 meV.

We set $\Delta_0=$250 meV, which induces a large gap between the conduction and valence bands of \ch{WSe2}. To model the low energy physics of graphene, we choose valley-specific Hamiltonian of the following form,
\begin{equation} \label{eq_3}
	H_k^{g}= v_F^{g}[\xi \sigma_x k_x+\sigma_yk_y]~.
\end{equation}
Here, $v_F^g$=3.46 eV.\AA ~is the Fermi velocity of graphene. Eqn.~\ref {eq_3} represents a gapless Dirac dispersion for the graphene sector. The coupling between the two layers is captured by the generalized hopping

\begin{equation}
	H_t=
	\begin{pmatrix}
		t_d & t_o \\
		t_o^* & t_d
	\end{pmatrix} \sigma_x~,
\end{equation}%
where $t_o$ = $t_r - i\xi t_i$.

In a perfectly aligned graphene-\ch{WSe2} heterostructure, the off-diagonal term $t_o$ is expected to be zero. However, in realistic experimental scenarios, misalignment of the heterostructure, lattice mismatch-induced strain, and small angular mismatch can result in a non-zero off-diagonal hopping, which can be complex in general~\cite{ho2021hall, Sinha2021, Chakraborty_2022,Chakraborty2022T}. In our calculation, we set the hopping strength $t_{d}=$50~meV and $t_r=t_i=$ 10~meV. The proximity effect of the \ch{WSe2} layer opens up a gap at the Dirac crossing of the graphene bands. The induced band gap of graphene gets enhanced with increased hopping strength (primarily $t_d$). The off-diagonal terms of the generalized hopping matrix induce an asymmetry in the low energy band dispersion (see Fig.\ref{Fig:contour}) in the vicinity of the band edges. More importantly, the axis of band structure asymmetry in the conduction band and the valance band is different (almost orthogonal). This band structure is reminiscent of that seen in strained bilayer graphene ~\cite{ho2021hall}.

Such an asymmetric band structure in the conduction and valance band  is analogous to a tilted massive Dirac dispersion, with the tilt direction being orthogonal for the conduction band and the valance band. We have plotted the fixed energy contours near the conduction and valence band edges in Fig.~\ref{Fig:contour}.  We have explicitly checked that the time-reversal symmetry is preserved and the energy dispersion relation satisfies $E_k^{K}= E_{-k}^{K^\prime}$. The symmetric response calculated from the composite graphene-\ch{WSe2} model suggests the Drude Hall resistivity $\sigma_{xy}^D$ changes sign when we go from the valence to conduction sector, similar to what we obtained in our experimental results.
As an additional check, we have also calculated the linear anomalous Hall response of the system, which turns out to be zero (within bounds of numerical accuracy) in presence of time reversal symmetry (see Fig.~\ref{Fig:sigma}).

The effect of the external perpendicular electric field is introduced by adding a diagonal Hamiltonian.
\begin{equation}
	H_\perp =
	\begin{pmatrix}
		\Delta \mathbb{I} & 0 \\
		0 & - \Delta \mathbb{I}
	\end{pmatrix}~.
\end{equation}
Figure~\ref{fig:s1} shows the band dispersion evolution with a perpendicular electric field. The band dispersion essentially undergoes an insulator-to-metal transition with the electric field (see Fig.~\ref{fig:s1}).

\section{II. Generic low energy model for sign reversing symmetric Hall effect}
The observed zero field Hall effect in our graphene-\ch{WSe2} devices has two important features: i)~it is symmetric in nature, and ii)~it changes sign across the charge neutrality point. To understand these, we note that in crystals with reduced symmetries, an anisotropic band dispersion can give rise to a symmetric Hall conductivity with $\sigma_{xy}(B=0)=\sigma_{yx}(B=0)$. This is typically interpreted as the Drude Hall conductivity or as the pseudo-planar Hall conductivity~\cite{ho2021hall}. An elegant way to see this is by considering a system with an anisotropic longitudinal Drude conductivity $\sigma'_{xx} \neq \sigma'_{yy}$, for a given set of coordinates (say $x'-y'$).  This anisotropic conductivity can be induced either by strain, heterostructure misalignment, or small moir\'{e} angle inhomogeneity. Rotating this conductivity matrix ($2 \times 2$ for a 2D system) to find the conductivity matrix in the device coordinate system (say $x-y$), we have $\sigma=R(-\theta) \sigma' R(\theta)$. As long as this rotation is not by integer multiples of $90^\circ$, there will always be a finite $\sigma_{xy}$ in the device coordinate system.
%
%

To show the existence of the symmetric Drude Hall effect in graphene-based devices, let us consider the case of gapped graphene with a tilted dispersion. It can be described by the following two-band model Hamiltonian for each valley (denoted by $\xi = \pm 1$),
\begin{equation} \label{ham_tilt}
	{\mathcal H}^\xi = v_F (\xi \sigma_x k_x + \sigma_y k_y) + \Delta \sigma_z  +  \xi \sigma_0(v_{ty}k_y + v_{tx} k_x)~.
\end{equation}
Here, the symbols have their usual meaning. The second term induces gap-opening between the bands, and the third term causes opposite orientations of the tilt in the two valleys. The opposite tilt in the two valleys preserves time-reversal symmetry. The tilt in each valley of the Hamiltonian in Eq.~\eqref{ham_tilt} breaks particle-hole symmetry. For this tilted massive Dirac Hamiltonian, we can calculate the symmetric Drude Hall conductivity to be \cite{Ashcroft76,ho2021hall}
\begin{equation} \label{DrudeH}
	\sigma^{\rm D}_{xy}=  e^2 \tau \sum_{m,\xi} \int \frac{ dk_x dk_y} {4\pi^2}  v_{x}^{m,\xi} v_{y}^{m,\xi} \left(\frac{-\partial f}{\partial \epsilon}\right)_{\epsilon=\epsilon_m(k)} ~\approx~
	\frac{e^2 \tau}{ (2\pi)^2 \hbar^2} v_{tx} v_{ty} |\mu |~.
\end{equation}
Here, $m$ denotes the sum over bands, $\xi$ is the valley index to be summed over, and $f (\epsilon)$ is the Fermi function.
Note that in deriving the second part of Eq.~\eqref{DrudeH}, we have retained only the lowest order contribution in $v_{tx}/v_F$ and $v_{ty}/v_F$ after systematically expanding the $v_x$, $v_y$ and the Fermi function up to second order contributions, and taken the $T \to 0$ limit. This establishes that graphene-based systems with anisotropic bands structure can support a symmetric Hall effect. Such an anisotropy can be induced by strain, substrate effects, or by angle misalignment in the case of heterostructures such as graphene-\ch{WSe2}.
We note the following here: i) Without tilt (or band anisotropy), the Drude mechanism can not generate a symmetric Hall conductivity in massive graphene. ii) We obtain the same Drude Hall conductivity for both valleys with opposite tilt orientation. iii) Most importantly, the Hall conductivity generated in this manner does not change its sign as we move from the valence to the conduction band. However, in our experiment, the Hall conductivity changes sign across the charge neutrality point. This mandates us to go beyond the tilted Dirac model described by Eq.~\eqref{ham_tilt}.

To understand the origin of sign reversing symmetric $\sigma_{xy}$  from Eq.~\eqref{DrudeH}, we note that it can arise if either of $v_{tx}$ or $v_{ty}$ have a different sign in the conduction band and the valance band. As an example, this can be achieved by considering the following set of band dispersion
\begin{eqnarray} \label{anisotropic_dis1}
	\epsilon_{k}^c &=& v_{tx} k_x + v_{ty} k_y + \sqrt{ v_F^2(k_x^2+k_y^2) + \Delta^2 }
	\\
	\epsilon_{k}^v &=& v_{tx} k_x - v_{ty} k_y + \sqrt{ v_F^2(k_x^2+k_y^2) + \Delta^2 }~.\label{anisotropic_dis2}
\end{eqnarray}
Here, the superscript $c~(v)$ refers to the conduction (valence) band.
These band dispersions have orthogonal tilts or orthogonal anisotropy directions in the conduction and valence bands.
For these set of bands, we have $\sigma_{xy}(\mu > \Delta) = -\sigma_{xy} (\mu<-\Delta) = \frac{e^2 \tau}{ (2\pi)^2 \hbar^2} v_{tx}v_{ty} |\mu |~.$

These simple calculations show that i) the anisotropy of the electronic dispersion can give rise to a symmetric Hall effect (or the Drude Hall effect) and, more importantly, ii) the sign reversal can arise from the conduction band and valance bands being anisotropic along different axes.
The scenario in our experiment is similar. The graphene-\ch{WSe2} band structure has an anisotropy, which is similar to the example presented above. Specifically, the anisotropy axis in the conduction band and the valence band are orthogonal to each other (see Fig.~\ref{Fig:contour}) and mimic the setup above with orthogonal tilt directions in the conduction and valance bands as described by Eq.~\eqref{anisotropic_dis1}-\eqref{anisotropic_dis2}. Consequently, the measured Drude Hall conductivity is opposite when we move from the valence to the conduction band.

\section{III. Drude longitudinal conductivity}
We calculate $\sigma_{xx}$ using the expression of the Drude conductivity\cite{Ashcroft76}
\begin{equation}
	\sigma_{xx}= e^2 \tau \sum_{m,\xi} \int \int \frac{ dk_x dk_y} {4\pi^2}  v_{x}^{m,\xi} v_{x}^{m,\xi} (\frac{-\partial f}{\partial \epsilon})_{\epsilon=\epsilon_m(k)}~.
\end{equation}
The band velocity is defined as $\hbar v_x^{m,\xi}=\partial \epsilon_{m,\xi}/ \partial k_x$, where $m$ is the band index. The longitudinal conductivity ($\sigma_{xx}$), which follows the density of states (DOS), shows a $ W$-like pattern with an increase in the electric field. The calculated $\sigma_{xx}$ captured the qualitative nature of the inverse of the experimental resistivity (${\rm R}_{xx}$) plot of Fig.~4(a) of the main manuscript. The pseudo gap within the first and second valence (conduction) bands promotes the low conducting dips below (above) the Fermi energy, whereas for a finite electric field, the substantial DOS at Fermi energy promotes the metallic nature indicated by a peak at the $\sigma_{xx}$ of Fig.~4(c) of the main manuscript.

\begin{figure*}[t]
	\begin{center}
		\includegraphics[width=\columnwidth]{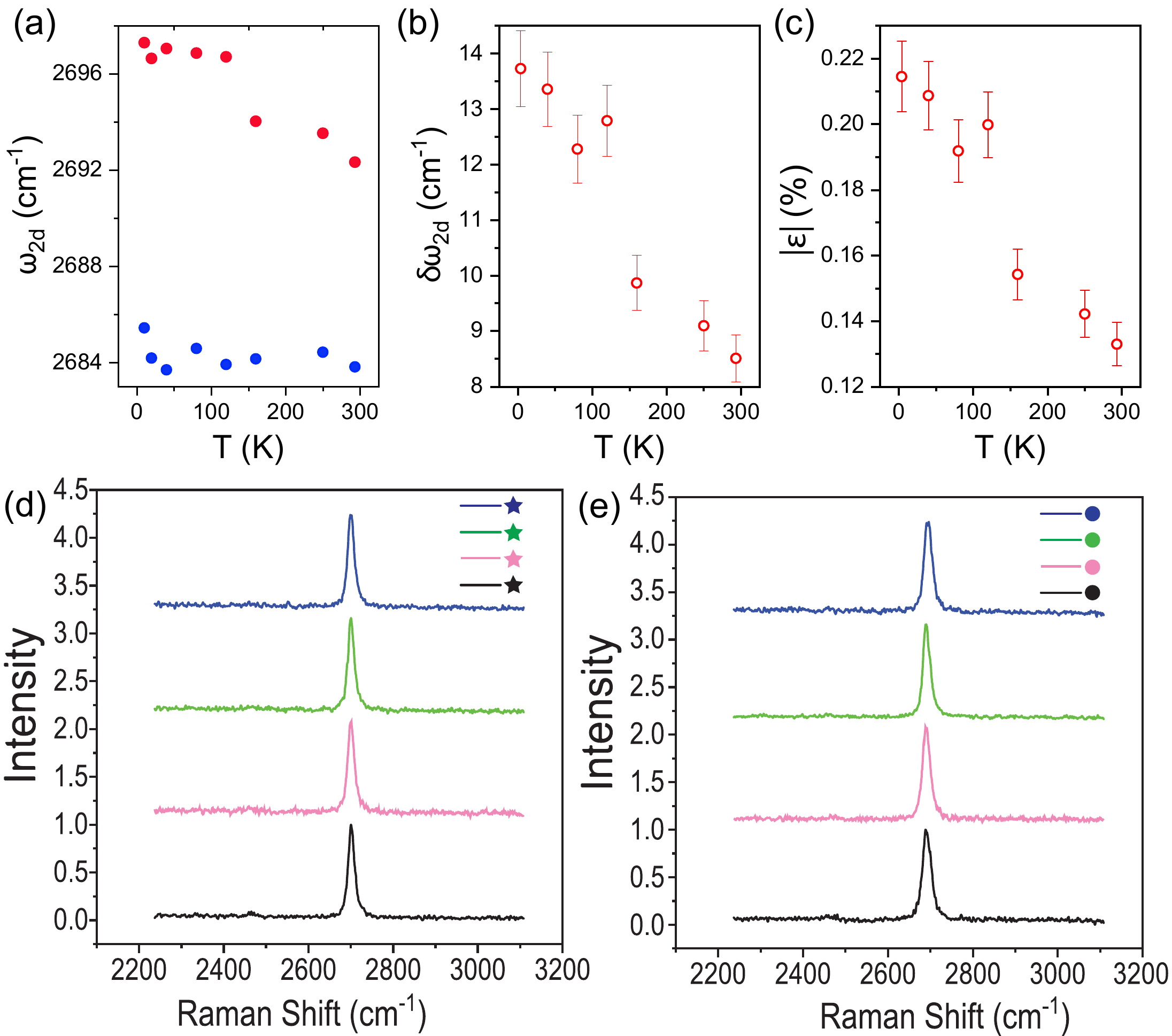}
		\caption{\textbf{Raman shift in the 2D band of graphene}  (a) Temperature variation of the measured Raman shift of the 2D peak of SLG (blue filled circles) and of SLG on  single-layer \ch{WSe2} (red filled circles).  (b) Plot of $\delta\omega_{2D}$ versus $T$ for SLG  on  single-layer \ch{WSe2}. (c) Plot of the $T$- dependence of the magnitude of the strain $|\epsilon|$ in SLG on single-layer \ch{WSe2}. 
			(d) Plots of Raman intensity versus Raman shift in hBN/SLG/\ch{WSe2}/hBN. Four data points are from four different locations. (e) Plots of Raman intensity versus Raman shift in the hBN/SLG/hBN region of the device. The data were taken at the four different locations in the sample. \label{Fig:S3}}
	\end{center}
\end{figure*}


\section{IV. Device fabrication}
Thin flakes of \ch{WSe2}, hBN, and single-layer graphene (SLG) were mechanically exfoliated on $\ch{Si/SiO_2}$ substrates. The thickness of the flakes was initially estimated from the color contrast under an optical microscope and later confirmed using Raman spectroscopy. This was followed by sequential pickup of each flake using Polycarbonate (PC) film at $\mathrm{90^{\circ}C}$. The assembled heterostructure was transferred on a new $\ch{Si/SiO_2}$ substrate~\cite{pizzocchero2016hot,wang2013one}. The heterostructure is then cleaned in chloroform, acetone, and IPA to remove the PC residue. The heterostructure was then annealed at $\mathrm{250^{\circ}C}$ for 3 hours. Electron beam lithography was used to define the contact and top gate electrodes. We used reactive ion etching (mixture of \ch{CHF3} and \ch{O2} gas) to etch top hBN to make one-dimensional edge contacts to graphene~\cite{Wang614}. For making the electrical contacts, Cr/Au (5~nm/60~nm) was deposited, followed by liftoff in hot acetone and cleaning in IPA. The unwanted hBN and graphene were removed  using E-beam lithography and dry etching to define the Hall bar. We transferred an hBN top of the device and fabricated a metallic top gate using lithography and thermal deposition.

\section{V. Raman shift and strain}
We used low-temperature  Raman spectroscopy in graphene \ch{WSe2} stack to estimate the strain in graphene.  High-quality SLG has two prominent Raman active modes, the G-mode (1580 $\mathrm {cm^{-1}})$ and the 2D-mode (2690 $\mathrm {cm^{-1}})$. In the presence of a uniaxial strain $\epsilon$, the shift in 2D peak has been measured to be  $\delta\omega_{2D}^{SLG}/\epsilon \sim  -64 \mathrm {cm^{-1}/\%} $~\cite{PhysRevB.79.205433}. Fig.~\ref{Fig:S3}(a) shows a comparison of the temperature-dependence of the Raman shift of the 2D band measured for graphene $\omega_{2D}^{SLG}$ and for graphene on \ch{WSe2} $\omega_{2D}^{SLG/WSe_2}$. In Fig.~\ref{Fig:S3}(b), we show a plot of the $\mathrm{T}$-dependence of $\delta\omega_{2D} =\omega_{2D}^{SLG/WSe_2} - \omega_{2D}^{SLG} $. One can see that the difference in the Raman shift of the 2D peak increases rapidly with a decrease in $T$; the positive value of $\delta\omega_{2D}$ indicates that the strain is compressive. The temperature dependence of the strain in graphene was extracted from the data in Fig.~\ref{Fig:S3}(b); its magnitude is plotted in Fig.~\ref{Fig:S3}(c). The data shows that SLG on single layer \ch{WSe2} undergoes a significant compressive strain of about 0.2\% at 4~K.


To check the charge and strain homogeneity in our devices,  we measured the Raman spectra of the comparative device consisting of hBN/SLG/hBN and hBN/SLG/\ch{WSe2}/hBN (these devices have common hBN and SLG). We measured the Raman spectra at several places in both devices. In Fig.~\ref{Fig:S3}(d), we show the Raman shift of hBN/SLG/\ch{WSe2}/hBN device at four representative points. In Fig.~\ref{Fig:S3}(e), we show the Raman shift of the hBN/SLG/hBN device at four representative points. We observed uniform intensity and no peak shift in these two individual regions attesting to the charge and strain homogeneity of the stacks. The spectra in each region are identical, again confirming the homogeneity of the device.

\section{VI. Additional Data for symmetric Hall effect on device SW1}

\begin{figure*}[t]
\begin{center}
	\includegraphics[width=\columnwidth]{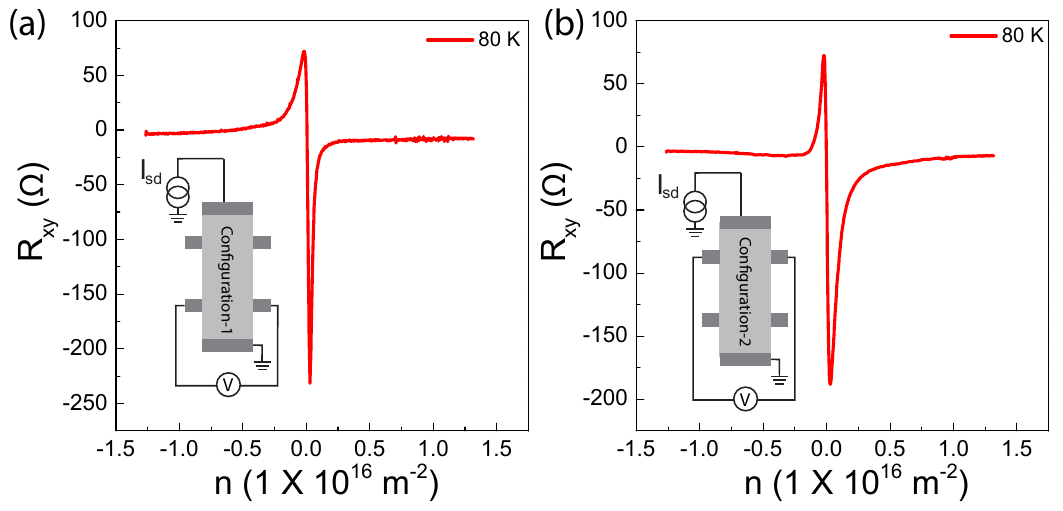}
	\caption{\small{\textbf{Data on device SW1.} Plot of transverse resistance versus number density in two different configurations for device SW1. The measurement in both configurations shows a sign reversing symmetric Hall effect.}}
	\label{Fig:S1_2}
\end{center}
\end{figure*}

The $\mathrm{R_{xy}}$ data from the device named SW1 measured across two pairs of transverse leads are compared in Fig.~\ref{Fig:S1_2} -- the data in both cases show a finite zero-magnetic-field transverse signal with sign depending on whether the charge carriers are electrons or holes. The $n_0$ for this device was estimated to be $2\times10^{14}$~m$^{-2}$ from the measured longitudinal resistance. The $\mathrm{R_{xy}}(B=0)$ changes sign for over range of $n\sim \pm 4\times 10^{15}$~m$^{-2}$ which is much larger than the $n_0$. So the study reported in this manuscript is outside the range of electron-hole puddles and, we believe,  unaffected by it.

\begin{figure*}[t]
\begin{center}
	\includegraphics[width=\columnwidth]{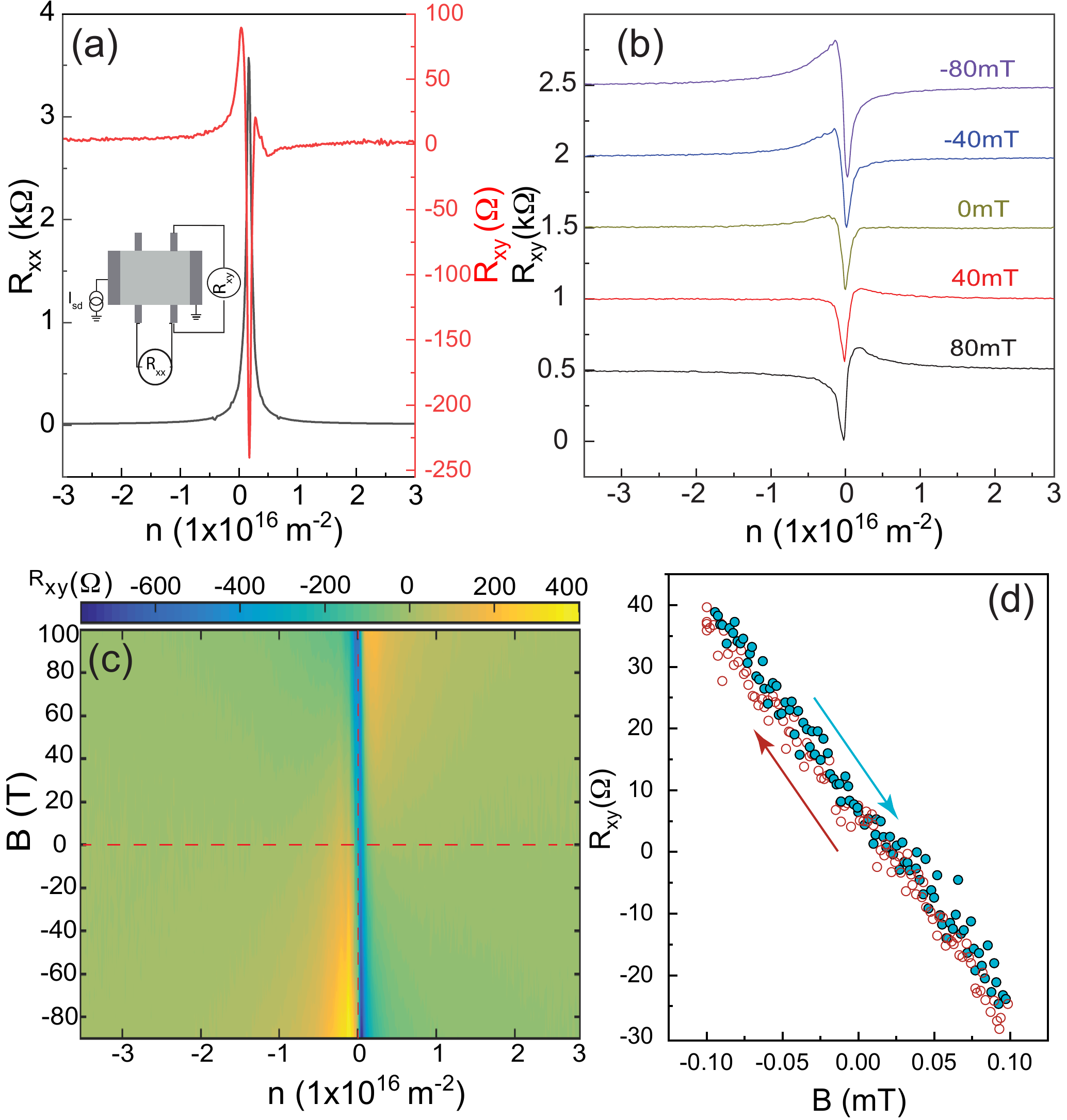}
	\caption{\small{\textbf{Data on device SW2.} (a) Plot of longitudinal and transverse resistivity versus number density for device SW2. (b) Plot of $\mathrm{R_{xy}}$ at small magnetic field values measured for device SW2. (c) A 2D map of the transverse resistance $\mathrm{R_{xy}(B)}$ in the $n - B$ plane; the data shows a finite Hall signal at $\mathrm{B=0}$~T. (d) Plot of magnetoresistance in a small magnetic field at $D=-0.3$~V/nm displacement field. The data were taken at $n=-2\times10^{16}\mathrm{m^{-2}} $. The red open circles are the data for $B$-sweep from positive to negative values; the filled circles are the data for $B$-sweep from negative to positive values. The arrows indicate the direction of the magnetic field sweep.}}
	\label{Fig:S1}
\end{center}
\end{figure*}


\section{VII. Additional Data for symmetric Hall effect on device SW2}

Fig.~\ref{Fig:S1}(a) shows the data for zero-field longitudinal and transverse resistance in device SW2; one can see the appearance of a finite $\mathrm{R_{xy}(B=0)}$ that changes its sign near the Dirac point. Interestingly, the $\mathrm{R_{xy}(B=0)}$ seems to have contribution from $\mathrm{R_{xx}(B=0)}$, probably due to misaligned probes. The Fig.~2(e) of the main text presents the $\mathrm{B=0}$ transverse signal measured in two different configurations, configuration~1 measures $\mathrm{R_{xy}(B=0)}$ while configuration~2 measures $\mathrm{R_{yx}(B=0)}$. The two signals overlap exactly
with each other. Note that this is one expects from the Onsager relation $\mathrm{R_{xy}(B)} = \mathrm{R_{xy}(-B)}$ for $\mathrm{B=0}$.


Fig.~\ref{Fig:S1}(b) shows the line plots of the transverse signal measured in device SW2 in the presence of a small perpendicular magnetic field.
The data show smooth evolution of the time-reversal symmetric Hall effect (at B=0) into the classical Hall signal (to $\mathrm{B\neq0}$). This can be better appreciated from Fig.~\ref{Fig:S1}(c), which is a 2D map of the transverse signal in the $n$-$\mathrm{B}$ plane.

The measured magnetoresistance in our devices is non-hysteretic (Fig.~\ref{Fig:S1}(d)). This is clear evidence of the absence of ferromagnetism in the system.

\section{VIII. Transverse resistance Data on hBN/graphene/hBN device at B=0}

\begin{figure}[t]
\includegraphics[width=\columnwidth]{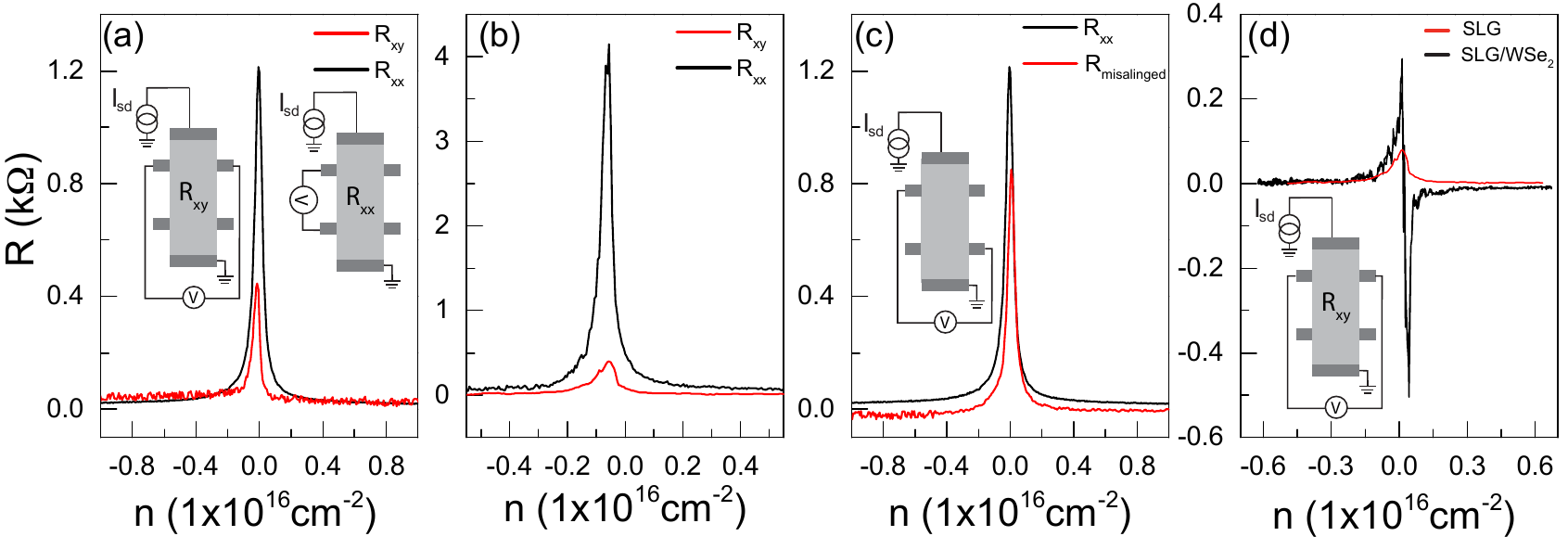}
\small{\caption{\textbf{Comparison of $\mathrm{B=0}$ transverse resistance in SLG and SLG/\ch{WSe2} devices.} Plots of measured $\mathrm{R_{xx}(B=0)}$ (black line) and $\mathrm{R_{xy}(B=0)}$ (red line) versus number density for (a) device SG3 and  (b) device SG4. The insets in (a) show the measurement configurations common to both panels. (c) Plots of measured $\mathrm{R_{xx}(B=0)}$ (black line)  and $\mathrm{R_{xy}(B=0)}$ with intentionally misaligned transverse probes (red line)  versus number density for device SG4. The measurement configuration is shown in the inset. (d) Comparison of measured $\mathrm{R_{xy}(B=0)}$ in bare SLG device (red line) and SLG/\ch{WSe2} heterostructure device (black line). The data for the SLG device has the same sign for electrons and holes while that for the SLG/\ch{WSe2} changes sign as the Fermi energy moves from the conduction band (positive $n$) to the valence band (negative $n$).}
	\label{Fig:S6}}
\end{figure}

To complement the data presented in the main manuscript, this section presents the data of zero-magnetic field transverse resistance in two hBN/graphene/hBN devices -- SG3 and SG4. The data have been plotted in Fig.~\ref{Fig:S6}(a) and (b). In both plots,  the transverse resistance $\mathrm{R_{xy}(B=0)}$ is plotted in red, and the longitudinal resistance $\mathrm{R_{xx}(B=0)}$ is plotted in black. In both devices, we find a finite $\mathrm{R_{xy}(B=0)}$ around the charge neutrality point. This has been seen previously by several groups and is usually attributed to misaligned transverse voltage probes or to impurity scattering.  We have also measured the $\mathrm{R_{xy}(B=0)}$ in a configuration where the probes measuring the transverse voltage are intentionally misaligned by a large amount (see Fig.~\ref{Fig:S6}(c)). In all these cases, the $\mathrm{R_{xy}(B=0)}$ mimics the $\mathrm{R_{xx}(B=0)}$ (with a diminished amplitude) signal; it does not change sign at the charge neutrality point.

Thus, there is a very important distinction between the $\mathrm{R_{xy}(B=0)}$ observed in bare graphene and the observations on strained SLG/\ch{WSe2} reported in the main manuscript. In the data presented in Fig.~\ref{Fig:S6}(a-c), the measured $\mathrm{R_{xy}(B=0)}$ has the same sign for both types of carriers. This is in stark contrast to the $\mathrm{R_{xy}(B=0)}$ in  SLG/\ch{WSe2}  devices where the $\mathrm{R_{xy}(B=0)}$ measured for electrons and holes have opposite signs (Fig.~\ref{Fig:S6}(d)).

\providecommand{\latin}[1]{#1}
\makeatletter
\providecommand{\doi}
{\begingroup\let\do\@makeother\dospecials
	\catcode`\{=1 \catcode`\}=2 \doi@aux}
\providecommand{\doi@aux}[1]{\endgroup\texttt{#1}}
\makeatother
\providecommand*\mcitethebibliography{\thebibliography}
\csname @ifundefined\endcsname{endmcitethebibliography}
{\let\endmcitethebibliography\endthebibliography}{}

\end{document}